\documentclass[11pt,a4paper]{article}

\usepackage[margin=2.6cm]{geometry}
\usepackage{amsmath,amssymb,amsthm}
\usepackage{booktabs}
\usepackage{graphicx}
\usepackage[hidelinks]{hyperref}
\usepackage{microtype}
\usepackage{setspace}
\usepackage{caption}
\usepackage{enumitem}

\setlist{nosep,leftmargin=*}
\theoremstyle{plain}
\newtheorem{conjecture}{Conjecture}

\title{An Empirical Square-Root Law for the Degree Discrepancy\\
of Binary $m$-Sequences}
\author{Joonas P\"a\"akk\"onen\\
Dalarna University, Borl\"ange, Sweden\\
\texttt{jpa@du.se}}
\date{November 11, 2026}

\begin{document}
\maketitle

\begin{abstract}
Let $k\ge 2$, $N=2^k$, $M=N-1$ and let $\Delta(r)$ be the range,
over all starting offsets, of the total-variation degree discrepancy of
a length-$r$ window of a primitive binary linear feedback shift register
(LFSR) of period $M$. Degree-discrepancy profiles for every $k\le 28$
support the scaling $\max_r\Delta(r)=\Theta(\sqrt{N})=\Theta(2^{k/2})$.
The tabulated peaks are maxima on a discrete grid, hence lower bounds on
the exhaustive maximum. The normalized peaks
$d_k=\max_{r\in R_k}\Delta(r)/2^{k/2}$
oscillate between values near $1/2$ and values that approach, but do
not exceed, $1$. The data suggest $\liminf d_k=1/2$ and $\limsup d_k\le 1$.
We conjecture the same square-root law for the exhaustive maximum,
with those normalized envelopes. A Brownian-bridge heuristic accounts for the
square-root \emph{scale}. The same profiles do not place the recorded
maximizer at mid-path: after folding by the complement identity
$\Delta(M-r)=\Delta(r)$, the median of $\min(\rho_k^\ast,1-\rho_k^\ast)$
is $0.35$, and several $k$ (notably $16$ and $28$) peak on a short window.
Mid-horizon $\Delta$ remains a large fraction of the peak, so the square-root
scale can still come from a Brownian-bridge envelope. The same range has two
readings of the starting offset: a pairing fragment in serial-knockout scheduling,
and a change to the plaintext on an LFSR orbit. The location of the maximizer
is not part of the conjecture. Open questions that separate the
conjecture from a theorem are listed.
\end{abstract}



\section{Introduction}
\label{sec:intro}

Let $k\ge 2$ be an integer and let $(s_t)_{t=0}^{2^k-2}$ be a binary
maximal-period sequence ($m$-sequence) of period $2^k-1$ generated by a
primitive linear feedback shift register (LFSR) of length $k$.
Equivalently, the classical trace sequence $s_t=\mathrm{Tr}(\alpha^t)$
records successive powers of a primitive element $\alpha\in\mathbb{F}_{2^k}$.
In the present work we work with the associated \emph{degree sequence}:
if the LFSR state at time $t$ is viewed as an integer in
$\{1,2,\dots,2^k-1\}$, its degree is the position of the highest set bit,
\[
\deg(t)=\lfloor\log_2(\text{state at time $t$})\rfloor
\in\{0,1,\dots,k-1\}.
\]
(The zero state never occurs in a maximal-period register.) Because the
orbit visits every nonzero state exactly once, the induced degrees follow
the highest-bit weights of $\{1,\dots,2^k-1\}$: the value $j$ appears
exactly $2^j$ times in every full period.

In earlier work~\cite{Paakkonen2026DAM} the degree discrepancy of
contiguous windows of a fixed length $r$ was studied from an optimization
viewpoint: one seeks the starting position that makes the empirical
degree histogram as close as possible to the ideal frequencies. The
present note addresses the complementary question. We examine how large
the gap between the best and the worst window of a given length $r$ can
become, and we do so uniformly in the horizon $r$. Write $\Delta(r)$ for
that gap (the precise definition is given in Section~\ref{sec:defs}).
Extensive computation of degree discrepancy profiles for all $k\le 28$
supports the square-root law
\[
\max_{1\le r\le 2^k-1}\Delta(r)=\Theta\bigl(2^{k/2}\bigr)
\]
or, equivalently,
\[
\max_{1\le r\le 2^k-1}\Delta(r)=\Theta(\sqrt{N})
\]
with $ N=2^k $,
which is why we call the relation a square-root law.
Each profile evaluates $\Delta(r)$ only on a discrete subset $R_k$ of
window lengths. The observed profile maxima already exhibit the
square-root growth; because a profile maximum is a lower bound on the
exhaustive maximum, the same growth is expected for $\max_r\Delta(r)$
itself. For $k\le 4$ rounding already includes every window length, so
the tabulated peak is exhaustive; for $k\ge 5$ it is a lower bound. The
corresponding normalized peaks
\[
d_k=\frac{\max_{r\in R_k}\Delta(r)}{2^{k/2}}=\frac{\max_{r\in R_k}\Delta(r)}{\sqrt{N}}
\]
appear to satisfy
\begin{equation*}
\liminf_{k\to\infty} d_k=\frac12
\end{equation*}
and
\begin{equation*}
\limsup_{k\to\infty} d_k\le 1.
\end{equation*}
No computed $d_k$ exceeds $1$; the largest value in Table~\ref{tab:peaks}
is $d_{15}\approx 0.98$. Those grid envelopes motivate
Conjecture~\ref{conj:sqrt} for the exhaustive maximum. A simple
Brownian-bridge heuristic accounts for both the square-root scale and
the lower envelope $1/2$; the same heuristic suggests that the upper
value $1$ is a natural free-walk reference rather than a sharp
constant. We do not conjecture that the maximizer sits near
$\rho=1/2$.

The same profiles do \emph{not} place that maximum at mid-path. A
Brownian bridge has variance $\rho(1-\rho)$ and is therefore typically
largest near $\rho=r/M=1/2$. The first recorded maximizer
$\rho_k^\ast=r_k^\ast/M$ on $R_k$ wanders. After folding by the
complement identity $\Delta(M-r)=\Delta(r)$, the median of
$\min(\rho_k^\ast,1-\rho_k^\ast)$ is $0.35$, and several $k$---notably
$k=16$ and $k=28$---peak on a short window. Mid-horizon $\Delta$ remains
a large fraction of the recorded peak, so the continuum story can still
set the \emph{scale} of $\max_r\Delta(r)$ even when it does not locate
the maximizer. This distinction is the main experimental point beyond the scale
itself: the square-root comparison is a statement about height, not
about the horizon at which that height is attained.

The same quantities admit two readings of the starting offset. In a
serial-knockout competition the offset selects a pairing fragment of
the $m$-sequence. In the cryptographic reading the offset is a change
to the plaintext. The second is only a reading of the same statistic,
not a cipher analysis.

The paper is organized as follows. Section~\ref{sec:related} situates
the result in the literature. Section~\ref{sec:defs} fixes notation.
Section~\ref{sec:methods} describes the algorithms used in the
experiments. Section~\ref{sec:evidence} presents the numerical evidence
for the scale, and Section~\ref{sec:location} records where the
profile maxima sit. Section~\ref{sec:bridge} develops the
Brownian-bridge explanation and explains why it should be read as an
envelope. Section~\ref{sec:apps} discusses the scheduling interpretation
and a short cryptographic reading. Section~\ref{sec:open} states the
square-root law as a conjecture and lists the questions that separate
it from a theorem.

\section{Related work}
\label{sec:related}

The word ``discrepancy'' is used in several distinct senses in the
combinatorial and analytic literature. In the classical theory of
uniform distribution and geometric discrepancy (see, e.g.,
Matou\v{s}ek's monograph~\cite{Matousek1999}) one measures how evenly a
point set fills a continuous domain. A different, purely arithmetic
meaning appears in the Erd\H{o}s discrepancy problem, resolved by
Tao~\cite{Tao2016}: for any sequence $f:\mathbb{N}\to\{\pm 1\}$ the
partial sums along homogeneous arithmetic progressions are unbounded.
The quantity studied in the present note is closer in spirit to the
former tradition, yet is discrete and cyclic: it is the $\ell_1$-deviation
(and its range over starting positions) of an empirical degree histogram
from the ideal frequencies on a maximal-period binary $m$-sequence.
While the underlying algebraic object is multiplicative, the degree
discrepancy itself is a sliding-window statistic on a finite cycle
rather than an asymptotic statement about arithmetic progressions.

Binary $m$-sequences generated by primitive linear feedback shift
registers are classical objects in the theory of pseudorandom sequences;
their correlation and balance properties are treated comprehensively by
Golomb and Gong~\cite{GolombGong2005} and in the finite-field
literature~\cite{LidlNiederreiter1997}. Uniform-distribution results for
linear recurring sequences over finite fields go back at least to
Niederreiter--Shiue~\cite{NiederreiterShiue1977,Niederreiter1980} and
were later extended by Niederreiter--Xing~\cite{NiederreiterXing1996}
and Herendi~\cite{Herendi2004}. None of these sources examines the range
of the degree-histogram discrepancy as a function of window length,
which is the quantity $\Delta(r)$ studied here.

On the Erd\H{o}s-discrepancy side, recent work of Klurman, Mangerel and
Ter\"av\"ainen~\cite{KlurmanMangerelTeravainen2024} develops pretentious
and Fourier-analytic methods for multiplicative functions over short
intervals in function fields---a setting conceptually close to short
contiguous fragments of a multiplicative orbit in
$\mathbb{F}_{2^k}^\times$. Continuum heuristics that replace discrete
degree discrepancy processes by Brownian bridges appear in several
modern treatments of irregularities of distribution; the present note
isolates the concrete square-root scaling that such heuristics predict
for the degree process of an $m$-sequence, and records that the same
heuristics do \emph{not} correctly locate the recorded maximizer.

Algebraic constructions of regular and nearly-regular tournament
structures via cyclotomic classes in finite fields have been studied by
Satake~\cite{Satake2019,Satake2021,Satake2025}. The optimization problem
of finding a single window of fixed length with minimal degree
discrepancy was treated in earlier work of the
author~\cite{Paakkonen2026DAM}; the present note addresses the
complementary extremal question of the maximal gap between best and
worst windows.

On the applied side, serial-knockout tournament design has received
renewed attention in the operations-research literature. Recent
contributions include the stability analysis of Lambers, Pendavingh and
Spieksma~\cite{LambersPendavinghSpieksma2025} and the broader survey of
Devriesere, Csat\'o and Goossens~\cite{DevriesereCsatoGoossens2025}.
The empirical square-root law below quantifies a concrete fairness
sensitivity for any bracket generated from a contiguous fragment of an
$m$-sequence. The same offset is a change to the plaintext on the
multiplicative orbit; that is a reading of $\Delta(r)$, not a cipher
analysis~\cite{GolombGong2005,Paakkonen2026DAM}.

Related open problems of a similar flavor appear in several neighboring
areas. The combinatorial discrepancy of windows in de~Bruijn sequences
(including those generated by primitive LFSRs) has been studied
experimentally and asymptotically~\cite{CooperHeitsch2010,GabricSawada2022},
yet a generic construction whose discrepancy matches the typical
behavior of a random binary string of the same length remains open.
Classical questions of Golomb on the characterization of $m$-sequences
by the simultaneous span-$n$ and two-level autocorrelation
properties~\cite{Golomb2007}, as well as the influence of the arithmetic
structure of the period $2^k-1$, are still unresolved and resonate with
the observed oscillation of the normalized peaks $d_k$. In the
uniform-distribution theory of linear recurring sequences, sharper
phase-dependent or higher-order discrepancy estimates continue to
present challenges~\cite{Larcher2014}. Finally, the quantitative
fairness of serial-knockout brackets generated from contiguous algebraic
fragments---precisely the setting in which $\Delta(r)$ measures offset
sensitivity---has only recently been
formalized~\cite{LambersPendavinghSpieksma2025}, and controlling the
worst-case degree imbalance within stable finite-field constructions is
an open design question.

\section{Definitions}
\label{sec:defs}

Let $\alpha$ be a primitive element of $\mathbb{F}_{2^k}$ and let
$\mathrm{Tr}:\mathbb{F}_{2^k}\to\mathbb{F}_2$ be the absolute trace. The
associated $m$-sequence may be realized as
\[
s_t=\mathrm{Tr}(\alpha^t),\qquad t=0,1,\dots,2^k-2.
\]
We work with the equivalent degree sequence obtained from the binary
representation of the successive powers of $\alpha$. Concretely, if the
state of a primitive LFSR of length $k$ is viewed as an integer in
$\{1,2,\dots,2^k-1\}$, its degree is the position of the highest set
bit:
\[
\deg(t)=\lfloor\log_2(\text{state at time $t$})\rfloor.
\]
(The zero state never occurs in a maximal-period register.)

Write $M=2^k-1$. For a window length $r\in\{1,\dots,M\}$ and a starting
index $t$ write
\[
c_j(t,r)=\#\{0\le i<r:\deg(t+i)=j\}
\]
(indices modulo $M$). The ideal frequencies are
\[
q_j=\frac{2^j}{M},\qquad j=0,1,\dots,k-1.
\]
The total-variation degree discrepancy of the window is
\[
D(t,r)=\frac12\sum_{j=0}^{k-1}\bigl|c_j(t,r)-rq_j\bigr|,
\]
where the factor $1/2$ converts the $\ell_1$ distance into total-variation
distance. The central object of the paper is the range
\begin{equation*}
\label{eq:Delta}
\Delta(r)=\max_{0\le t<M}D(t,r)-\min_{0\le t<M}D(t,r).
\end{equation*}
When $r=M$ every window is a full period, so $D(t,M)\equiv 0$ and
$\Delta(M)=0$.

The counts on complementary windows are linked by the exact global
balance of the cycle. For $1\le r\le M-1$ the windows $[t,t+r)$ and
$[t+r,t+M)$ partition one period, hence
\[
c_j(t,r)+c_j(t+r,M-r)=2^j
\]
and
\[
c_j(t+r,M-r)-(M-r)q_j=-(c_j(t,r)-rq_j).
\]
It follows that $D(t+r,M-r)=D(t,r)$ and therefore
\begin{equation}
\label{eq:complement}
\Delta(M-r)=\Delta(r).
\end{equation}
In particular the identity~\eqref{eq:complement} implies that the map
$\rho\mapsto\Delta(\lfloor\rho M\rfloor)$ is approximately symmetric
about $\rho=1/2$: a recorded maximizer at relative length $\rho$ has a
counterpart of the same size at a relative length differing from
$1-\rho$ by at most $1/M$. The folded location
\[
\min(\rho,1-\rho)
\]
is therefore the natural test of proximity to mid-path. Finally we set
\[
d_k=\frac{\max_{r\in R_k}\Delta(r)}{2^{k/2}},
\]
where $R_k$ denotes the discrete set of window lengths examined in the
computational profiles described in Section~\ref{sec:methods}. Thus
$d_k$ is a normalized profile peak. It is a lower bound on the
normalized exhaustive maximum
$\max_{1\le r\le M}\Delta(r)/2^{k/2}$.

\section{Computational methods}
\label{sec:methods}

All experiments rely on two elementary procedures. A single primitive
polynomial of degree $k$ is used for each $k$. The polynomials are
recorded in the accompanying code. Those of degree $2\le k\le 16$ appear
in the table of Maurer~\cite{Maurer2013}; those of degree
$17\le k\le 28$ appear in the table of Moon~\cite{Moon2005}. A primitive
polynomial of degree $k$ over $\mathbb{F}_2$ generates an LFSR orbit of
period $M=2^k-1$. At startup the code checks that each recorded
polynomial has degree $k$.

\begin{center}
\fbox{\parbox{0.92\linewidth}{\small
\textbf{Algorithm 1} (Degree sequence of a primitive LFSR).\\
\textbf{Require:} primitive polynomial $p$ of degree $k$.\\[2pt]
$P\leftarrow 2^k-1$;\quad $s\leftarrow 1$.\\
\textbf{for} $t=0$ \textbf{to} $P-1$ \textbf{do}\\
\quad $\mathrm{deg}[t]\leftarrow\lfloor\log_2 s\rfloor$;\quad
$s\leftarrow 2s$;\\
\quad \textbf{if} the $k$-th bit of $s$ is set \textbf{then}
$s\leftarrow s\oplus p$.\\
\textbf{end for}\\
\textbf{return} $\mathrm{deg}[0..P-1]$.
}}
\end{center}

Exact sliding-window evaluation (Algorithm~2) was used for all profile
points when $k\le 14$. For $15\le k\le 28$ the range $\Delta(r)$ was
estimated by Monte-Carlo sampling of random starting positions without
replacement: $1200$ samples when $r<M/5$, and $600$ samples otherwise.
The Monte-Carlo estimator is conservative: it can only under-estimate
the true range. The reported peaks for $k\ge 15$ are therefore lower
bounds on the exact profile maxima. The unequal sample sizes imply that
short windows are probed more thoroughly than long ones; a recorded
short-window peak (notably at $k=28$) is not on a strictly equal footing
with a mid-horizon estimate.

\begin{center}
\fbox{\parbox{0.92\linewidth}{\small
\textbf{Algorithm 2} (Exact range $\Delta(r)$ by sliding window).\\[2pt]
Initialize the histogram of the window $[0,r)$ and compute its degree
discrepancy $D$.\\
$D_{\min}\leftarrow D_{\max}\leftarrow D$.\\
\textbf{for} $t=1$ \textbf{to} $2^k-2$ \textbf{do}\\
\quad Remove $\mathrm{deg}[t-1]$, insert
$\mathrm{deg}[(t-1+r)\bmod(2^k-1)]$, update $D$.\\
\quad $D_{\min}\leftarrow\min(D_{\min},D)$,\quad
$D_{\max}\leftarrow\max(D_{\max},D)$.\\
\textbf{end for}\\
\textbf{return} $D_{\max}-D_{\min}$.
}}
\end{center}

The profiles themselves consist of a geometric sequence of relative
lengths in the short-window regime (up to $\rho\approx 0.3$) together
with a linear grid covering the middle and long-window regimes, yielding
$28$ evaluation points after rounding to integer $r$. For small $k$
rounding collapses several relative lengths onto the same $r$, so the
number of distinct windows is then less than $28$. For $k\le 4$ this
collapse already covers every $r\in\{1,\dots,M\}$. For $k\ge 5$ the grid
is a proper subset, so the quantity reported as $\max_r\Delta(r)$ in
Table~\ref{tab:peaks} is a profile maximum and a lower bound on the
exhaustive maximum. The square-root comparison is insensitive to a
modest missed increment on the scale $2^{k/2}$, but for $k\ge 5$
Table~\ref{tab:peaks} should not be read as an exhaustive
$\max_r\Delta(r)$, even in the exactly computed regime $k\le 14$. The
same caveat applies, more strongly, to the recorded argmax
$r_k^\ast\in R_k$ discussed in Section~\ref{sec:location}: a coarse grid
can mis-locate a maximum even when it estimates the height well.

\section{Numerical evidence for the scale}
\label{sec:evidence}

\begin{figure}[!ht]
\centering
\includegraphics[width=\linewidth]{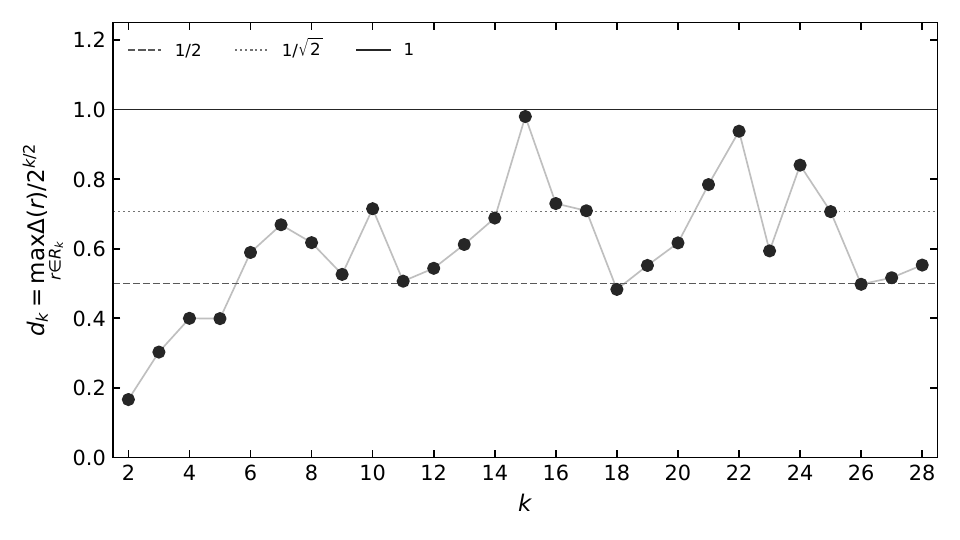}
\caption{Normalized peak degree discrepancy
$d_k=\max_{r\in R_k}\Delta(r)/2^{k/2}$ for $k=2,\dots,28$.
Horizontal lines mark the Brownian-bridge lower envelope $1/2$
(dashed), the intermediate scale $1/\sqrt{2}\approx 0.707$ (dotted),
and the free random-walk scale $1$ (solid).}
\label{fig:dk}
\end{figure}

Figure~\ref{fig:dk} shows the full sequence of normalized peaks $d_k$
for $2\le k\le 28$. Table~\ref{tab:peaks} records the corresponding peak
values.

Both the extreme short-window regime $\rho\to 0$ and the full-period
endpoint $\rho=1$ produce small values of $\Delta(r)$, as required
by~\eqref{eq:complement} and by the global balance of the cycle. Away
from those endpoints the profile is broader than a single bulk peak.
The first recorded maximizer $\rho_k^\ast=r_k^\ast/M$ on $R_k$
sometimes lies near $1/2$ (for instance $k=8,9,14,25$), sometimes in the
complementary long-window half (for instance $k=11,12,15,18,27$), and
occasionally on a comparatively short window ($k=16$ at
$\rho^\ast\approx 0.14$, $k=28$ at $\rho^\ast\approx 0.074$).
Because of~\eqref{eq:complement} a long-window recorded maximum is the
same size as its short-window complement, so the first-argmax convention
on a discrete grid does not by itself locate a unique physical regime.
In particular the data do not support a uniform statement that every
$k\ge 10$ attains its profile maximum inside $[0.25,0.55]$.
The same profiles do, however, give a stable height statement: even
when the first recorded maximizer sits on a wing, the mid-horizon value
of $\Delta$ remains a large fraction of the recorded peak
(Table~\ref{tab:argmax}).

The short-window recorded peak at $k=28$ should be read with the
sampling caveat of Section~\ref{sec:methods}: that point used $1200$
Monte-Carlo starts, whereas the mid-horizon grid used $600$. The largest
mid-horizon estimate on the same profile is still of the same order
($d_{28}$ would remain near $1/2$ if the short-window point were
discarded). The same remark applies, more mildly, to the short-window
first argmax at $k=16$. In both cases the location is an experimental
observation about the grid $R_k$, not part of Conjecture~\ref{conj:sqrt}.
A closer look at the sequence $(d_k)$ reveals a persistent oscillation
between the Brownian-bridge floor $1/2$ and the free-walk reference $1$.
No computed peak overshoots the line $1$. The largest normalized values
are $d_{15}\approx 0.98$ and $d_{22}\approx 0.94$; several later values
return to the floor, including $d_{26}\approx 0.50$, $d_{27}\approx 0.52$
and $d_{28}\approx 0.55$. The sampled value $d_{18}\approx 0.48$ sits
slightly below $1/2$, which is compatible with a conservative estimator
and should not be read as evidence that the liminf is strictly smaller
than $1/2$. The amplitude of the oscillation does not appear to decay
rapidly; whether the limsup is strictly less than $1$, equal to $1$, or
slowly diverging remains open (see Section~\ref{sec:open}).

One possible explanation of the oscillation is the arithmetic structure
of the period $M=2^k-1$: when $M$ is highly composite the degree process
might align with low-frequency characters for longer stretches, while a
Mersenne-prime period might mix more rapidly. The present range of $k$
contains too few Mersenne cases, and too strong a small-$k$ transient,
for this to be tested; a systematic comparison remains an open question.

\begin{table}[!ht]
\centering
\caption{Peak discrepancies obtained from the computational profiles
($k=2$ to $k=28$). The column $\max\Delta$ is the maximum of
$\Delta(r)$ on the $28$-point grid $R_k$. The next three columns are
the reference scales $c\cdot 2^{k/2}$, not normalized peaks; $d_k$ is
the dimensionless ratio $(\max\Delta)/2^{k/2}$. For $k\le 4$ the grid
already contains every $r$; for $k\ge 5$ the tabulated peak is a lower
bound on the exhaustive maximum. Entries for $k\le 14$ are exact on
the grid; entries for $k\ge 15$ are Monte-Carlo estimates and hence
lower bounds on the true profile maxima.}
\label{tab:peaks}
{\small
\setlength{\tabcolsep}{3.2pt}
\begin{tabular}{@{}rrrrrrrl@{}}
\toprule
&&&\multicolumn{3}{c}{scales $c\cdot 2^{k/2}$}&&\\
\cmidrule(lr){4-6}
$k$ & $M$ & $\max\Delta$ &
$c=\tfrac12$ & $c=\tfrac1{\sqrt{2}}$ & $c=1$ & $d_k$ & method \\
\midrule
 2 & 3 & 0.33 & 1.00 & 1.41 & 2.00 & 0.17 & exact \\
 3 & 7 & 0.86 & 1.41 & 2.00 & 2.83 & 0.30 & exact \\
 4 & 15 & 1.60 & 2.00 & 2.83 & 4.00 & 0.40 & exact \\
 5 & 31 & 2.26 & 2.83 & 4.00 & 5.66 & 0.40 & exact \\
 6 & 63 & 4.71 & 4.00 & 5.66 & 8.00 & 0.59 & exact \\
 7 & 127 & 7.57 & 5.66 & 8.00 & 11.31 & 0.67 & exact \\
 8 & 255 & 9.88 & 8.00 & 11.31 & 16.00 & 0.62 & exact \\
 9 & 511 & 11.91 & 11.31 & 16.00 & 22.63 & 0.53 & exact \\
10 & 1023 & 22.88 & 16.00 & 22.63 & 32.00 & 0.71 & exact \\
11 & 2047 & 22.93 & 22.63 & 32.00 & 45.25 & 0.51 & exact \\
12 & 4095 & 34.80 & 32.00 & 45.25 & 64.00 & 0.54 & exact \\
13 & 8191 & 55.41 & 45.25 & 64.00 & 90.51 & 0.61 & exact \\
14 & 16383 & 88.08 & 64.00 & 90.51 & 128.00 & 0.69 & exact \\
15 & 32767 & 177.37 & 90.51 & 128.00 & 181.02 & 0.98 & sampled \\
16 & 65535 & 186.83 & 128.00 & 181.02 & 256.00 & 0.73 & sampled \\
17 & 131071 & 256.70 & 181.02 & 256.00 & 362.04 & 0.71 & sampled \\
18 & 262143 & 247.34 & 256.00 & 362.04 & 512.00 & 0.48 & sampled \\
19 & 524287 & 399.62 & 362.04 & 512.00 & 724.08 & 0.55 & sampled \\
20 & 1048575 & 631.48 & 512.00 & 724.08 & 1024.00 & 0.62 & sampled \\
21 & 2097151 & 1135.72 & 724.08 & 1024.00 & 1448.15 & 0.78 & sampled \\
22 & 4194303 & 1920.25 & 1024.00 & 1448.15 & 2048.00 & 0.94 & sampled \\
23 & 8388607 & 1719.62 & 1448.15 & 2048.00 & 2896.31 & 0.59 & sampled \\
24 & 16777215 & 3441.20 & 2048.00 & 2896.31 & 4096.00 & 0.84 & sampled \\
25 & 33554431 & 4092.84 & 2896.31 & 4096.00 & 5792.62 & 0.71 & sampled \\
26 & 67108863 & 4078.11 & 4096.00 & 5792.62 & 8192.00 & 0.50 & sampled \\
27 & 134217727 & 5985.51 & 5792.62 & 8192.00 & 11585.24 & 0.52 & sampled \\
28 & 268435455 & 9060.46 & 8192.00 & 11585.24 & 16384.00 & 0.55 & sampled \\
\bottomrule
\end{tabular}
}
\end{table}

\section{Where the maximum sits}
\label{sec:location}

\begin{figure}[!ht]
\centering
\includegraphics[width=\linewidth]{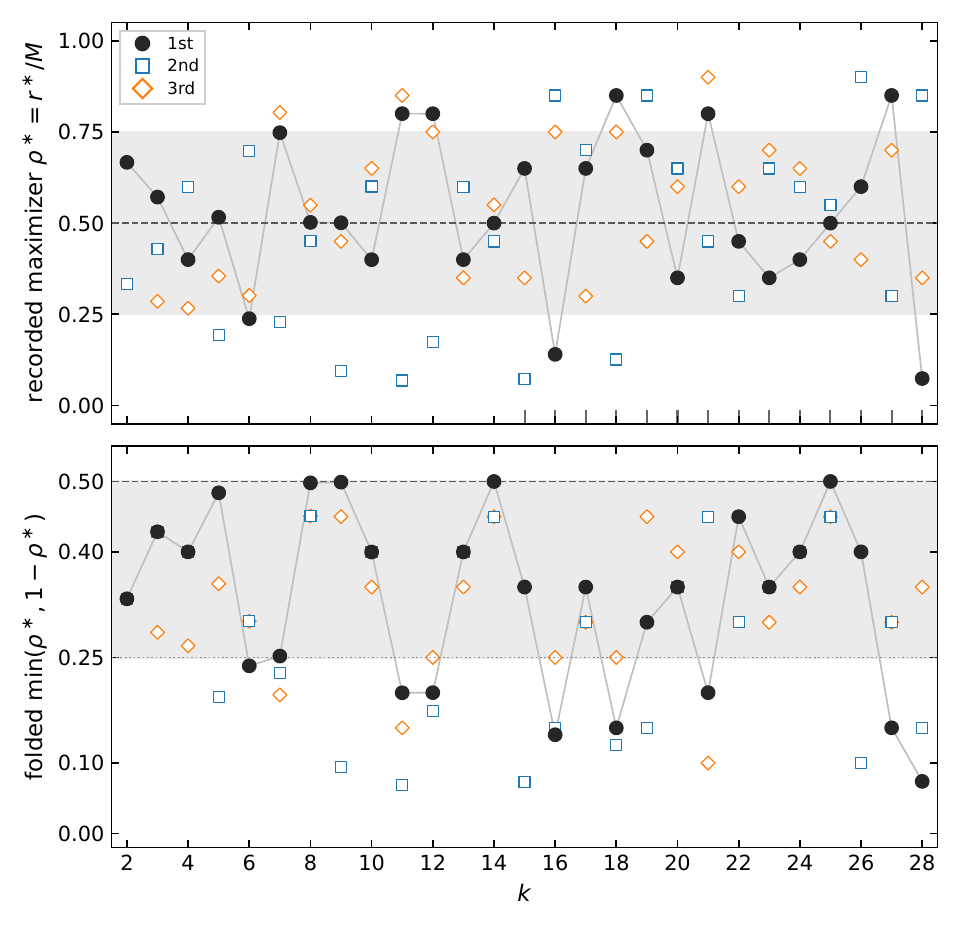}
\caption{Recorded maximizers of $\Delta(r)$ on the $28$-point grids
$R_k$. Top: the first three distinct window lengths, plotted as
$\rho=r/M$. Bottom: the same points after folding by
$\min(\rho,1-\rho)$, which is the mid-path test implied
by~\eqref{eq:complement}. The shaded band is the central half-cycle
$[1/4,3/4]$ (top) and its fold $[1/4,1/2]$ (bottom). Vertical ticks
mark sampled values ($k\ge 15$). The first argmax does not hover
around $1/2$.}
\label{fig:argmax}
\end{figure}

The square-root comparison of Section~\ref{sec:evidence} is a statement
about \emph{height}. It does not record the horizon at which that height
is attained. A Brownian-bridge heuristic for the scale has variance
$\rho(1-\rho)$ (Section~\ref{sec:bridge}), so if the continuum story
were also a prediction of location one would expect
$r_k^\ast/M$ to hover near $1/2$. Figure~\ref{fig:argmax} and
Table~\ref{tab:argmax} test that expectation on the existing grids
$R_k$, with no new LFSR scans.

Because of~\eqref{eq:complement}, a recorded maximizer at $\rho$ is
interchangeable with one at $1-\rho$. The folded coordinate
$\min(\rho,1-\rho)$ is therefore the quantity that tests ``near
mid-path''. On the $27$ values $k=2,\dots,28$ the first argmax does
\emph{not} hover around $1/2$:
\begin{itemize}[leftmargin=1.6em,itemsep=0.35em,topsep=0.5em]
  \item the folded location is below $1/4$ for eight values of $k$,
  including $k=16$ ($\rho^\ast\approx 0.14$) and $k=28$
  ($\rho^\ast\approx 0.074$);
  \item the median folded location is $0.35$, not $0.50$;
  \item the second and third distinct grid points are typically within
  ten percent of the peak and are not concentrated in the same third of
  the cycle.
\end{itemize}
The recorded maximum is therefore a plateau rather than a single bulk
spike. Evaluating the same profiles at the grid point nearest
$\rho=1/2$ still yields at least $61\%$ of the recorded peak (median
$82\%$). Thus the square-root comparison is not an artefact of a single
short-window outlier, even when that outlier is the recorded maximizer.

Two caveats already stated in Section~\ref{sec:methods} apply with extra
force to location. First, $r_k^\ast$ is an argmax on $R_k$, not an
exhaustive $\mathrm{argmax}_r$. A finer grid could move the recorded
horizon without changing the height on the scale $2^{k/2}$. Second, for
$k\ge 15$ the estimator is Monte Carlo, and short windows used twice as
many samples as mid-horizon windows. The short-window first argmax at
$k=28$ is the case in which that imbalance is most relevant. Even if
that point were discarded, Table~\ref{tab:argmax} would still contain
several folded locations at or below $0.20$ in the exact regime
($k=6,11,12$).

\begin{table}[!ht]
\centering
\small
\setlength{\tabcolsep}{3.6pt}
\caption{First recorded maximizer $\rho_k^\ast=r_k^\ast/M$ on $R_k$,
its complement-folded location, and the ratio of mid-horizon
$\Delta$ (grid point nearest $\rho=1/2$) to the recorded peak.
The last column lists the top three distinct grid points as
$\rho$~(fraction of the peak).}
\label{tab:argmax}
\begin{tabular}{@{}rrrrcl@{}}
\toprule
$k$ & method & $\rho^\ast$ & folded & $\Delta(\tfrac12)/\max\Delta$ &
top-3 $\rho$ ($\Delta/\max$) \\
\midrule
 2 & exact   & 0.667 & 0.333 & 1.00 & $0.667\,(1.00)$,\ $0.333\,(1.00)$ \\
 3 & exact   & 0.571 & 0.429 & 1.00 & $0.571\,(1.00)$,\ $0.429\,(1.00)$,\ $0.286\,(0.83)$ \\
 4 & exact   & 0.400 & 0.400 & 0.83 & $0.400\,(1.00)$,\ $0.600\,(1.00)$,\ $0.267\,(0.87)$ \\
 5 & exact   & 0.516 & 0.484 & 1.00 & $0.516\,(1.00)$,\ $0.194\,(0.87)$,\ $0.355\,(0.87)$ \\
 6 & exact   & 0.238 & 0.238 & 0.72 & $0.238\,(1.00)$,\ $0.698\,(0.93)$,\ $0.302\,(0.93)$ \\
 7 & exact   & 0.748 & 0.252 & 0.72 & $0.748\,(1.00)$,\ $0.228\,(0.92)$,\ $0.803\,(0.91)$ \\
 8 & exact   & 0.502 & 0.498 & 1.00 & $0.502\,(1.00)$,\ $0.451\,(0.96)$,\ $0.549\,(0.96)$ \\
 9 & exact   & 0.501 & 0.499 & 1.00 & $0.501\,(1.00)$,\ $0.094\,(0.91)$,\ $0.450\,(0.91)$ \\
10 & exact   & 0.400 & 0.400 & 0.84 & $0.400\,(1.00)$,\ $0.600\,(1.00)$,\ $0.650\,(0.92)$ \\
11 & exact   & 0.800 & 0.200 & 0.82 & $0.800\,(1.00)$,\ $0.068\,(1.00)$,\ $0.850\,(0.92)$ \\
12 & exact   & 0.800 & 0.200 & 0.62 & $0.800\,(1.00)$,\ $0.174\,(1.00)$,\ $0.750\,(0.89)$ \\
13 & exact   & 0.400 & 0.400 & 0.85 & $0.400\,(1.00)$,\ $0.600\,(1.00)$,\ $0.350\,(0.98)$ \\
14 & exact   & 0.500 & 0.500 & 1.00 & $0.500\,(1.00)$,\ $0.450\,(0.94)$,\ $0.550\,(0.94)$ \\
15 & sampled & 0.650 & 0.350 & 0.66 & $0.650\,(1.00)$,\ $0.073\,(0.97)$,\ $0.350\,(0.95)$ \\
16 & sampled & 0.140 & 0.140 & 0.69 & $0.140\,(1.00)$,\ $0.850\,(0.94)$,\ $0.750\,(0.91)$ \\
17 & sampled & 0.650 & 0.350 & 0.63 & $0.650\,(1.00)$,\ $0.700\,(0.92)$,\ $0.300\,(0.89)$ \\
18 & sampled & 0.850 & 0.150 & 0.74 & $0.850\,(1.00)$,\ $0.126\,(1.00)$,\ $0.750\,(0.95)$ \\
19 & sampled & 0.700 & 0.300 & 0.70 & $0.700\,(1.00)$,\ $0.850\,(0.98)$,\ $0.450\,(0.97)$ \\
20 & sampled & 0.350 & 0.350 & 0.71 & $0.350\,(1.00)$,\ $0.650\,(0.99)$,\ $0.600\,(0.90)$ \\
21 & sampled & 0.800 & 0.200 & 0.83 & $0.800\,(1.00)$,\ $0.450\,(1.00)$,\ $0.900\,(0.98)$ \\
22 & sampled & 0.450 & 0.450 & 0.86 & $0.450\,(1.00)$,\ $0.300\,(0.99)$,\ $0.600\,(0.98)$ \\
23 & sampled & 0.350 & 0.350 & 0.65 & $0.350\,(1.00)$,\ $0.650\,(1.00)$,\ $0.700\,(0.85)$ \\
24 & sampled & 0.400 & 0.400 & 0.75 & $0.400\,(1.00)$,\ $0.600\,(0.98)$,\ $0.650\,(0.94)$ \\
25 & sampled & 0.500 & 0.500 & 1.00 & $0.500\,(1.00)$,\ $0.550\,(0.94)$,\ $0.450\,(0.94)$ \\
26 & sampled & 0.600 & 0.400 & 0.94 & $0.600\,(1.00)$,\ $0.900\,(0.99)$,\ $0.400\,(0.96)$ \\
27 & sampled & 0.850 & 0.150 & 0.82 & $0.850\,(1.00)$,\ $0.300\,(0.97)$,\ $0.700\,(0.97)$ \\
28 & sampled & 0.074 & 0.074 & 0.68 & $0.074\,(1.00)$,\ $0.850\,(0.91)$,\ $0.350\,(0.89)$ \\
\bottomrule
\end{tabular}
\end{table}

\section{Brownian-bridge heuristic}
\label{sec:bridge}

We now give a continuum explanation for the observed square-root scale
and for the lower envelope $1/2$. The same picture also produces a
natural free-walk reference of size $2^{k/2}$ for the upper envelope.
What it does \emph{not} predict is the location of the maximizer:
variance $\rho(1-\rho)$ is largest at mid-path, but the recorded
argmaxes on $R_k$ do not concentrate there. The heuristic is therefore
an envelope for the height of $\max_r\Delta(r)$, not a claim about
$r_k^\ast/M$.

Let $r=\lfloor\rho M\rfloor$. For each degree $j\in\{0,1,\dots,k-1\}$ the centered count
\[
X_j(t)=c_j(t,r)-rq_j
\]
is a cyclic sliding-window process. Because the underlying degree
sequence is itself generated by a linear recurrence of full period, the
vector $X(t)=(X_0(t),\dots,X_{k-1}(t))$ behaves, for large $k$ and for
relative window length $\rho$ bounded away from $0$ and $1$,
like a multivariate random walk on the cycle of length~$M$. The walk is
constrained by the exact global balance
\[
\sum_{j=0}^{k-1}X_j(t)=0
\]
and by the fact that the total number of occurrences of each degree
over a full period is deterministic. These constraints turn the walk
into a bridge: after time $M$ it must return to the origin. The
complement identity~\eqref{eq:complement} is the discrete counterpart of
the elementary symmetry $\mathrm{Var}(B^\circ(\rho))=\mathrm{Var}(B^\circ(1-\rho))$,
and it explains why a recorded maximizer need not sit near $\rho=1/2$
even if that is where the bridge variance is largest.

A standard Brownian bridge $B^\circ$ on $[0,1]$ (the continuum limit of
a random walk conditioned to return to zero) satisfies
\[
\mathrm{Var}(B^\circ(\rho))=\rho(1-\rho).
\]
The maximum of the variance occurs at $\rho=1/2$ and equals $1/4$.
Consequently the typical size of a single coordinate $X_j$ at the
mid-point is of order
\[
2^{k/2}\cdot\tfrac12.
\]
Because the ideal frequencies $q_j$ decay geometrically, the $\ell_1$
sum that defines $D(t,r)$ is dominated by a constant number of large
bins; the single-coordinate Brownian-bridge scale therefore carries over
to the range $\Delta(r)$ without an extra $\sqrt{k}$ factor. In
particular one expects
\[
\max_r\Delta(r)=\Theta\bigl(2^{k/2}\bigr).
\]
Evaluating the bridge variance at its maximum yields the concrete lower
reference $\tfrac12\cdot 2^{k/2}$. The numerical data up to $k=28$ are
consistent with this value being asymptotically sharp from below:
\[
\liminf_{k\to\infty}\frac{\max_r\Delta(r)}{2^{k/2}}=\frac12.
\]

That calculation predicts the \emph{scale} of $\Delta(r)$ at mid-path,
not the location of $\max_r\Delta(r)$. Figure~\ref{fig:argmax} shows
that the recorded maximizer often lies well away from $\rho=1/2$. The
appropriate reading of the heuristic is therefore an envelope: a
mid-path bridge of height $\tfrac12\cdot 2^{k/2}$ already accounts for
most of the observed peak (Table~\ref{tab:argmax}: median ratio
$0.82$), while the remaining increment can sit on a short or long wing.
In particular the continuum story should not be taken as a prediction
that $r^\ast/M$ concentrates near $1/2$. If the exhaustive maximizer
were later shown to return to a bulk interval, that would restore a
location statement; the present grids do not support one.

A free (unconditioned) random walk of length $M$ would fluctuate on the
larger scale $2^{k/2}$. The same free-walk scale appears in the data as
a practical upper envelope: several of the computed peaks approach
$1\cdot 2^{k/2}$ (most closely $d_{15}\approx 0.98$), but none exceed it
and none grow visibly faster. This suggests the working upper bound in Conjecture~\ref{conj:sqrt},
\[
\limsup_{k\to\infty}\frac{\max_r\Delta(r)}{2^{k/2}}\le 1.
\]

The heuristic is of course not a proof. Three obstacles stand between
the continuum picture and a rigorous invariance principle:
\begin{itemize}[leftmargin=1.6em,itemsep=0.35em,topsep=0.5em]
  \item the degree sequence is deterministic and highly algebraically
  structured, not a sequence of independent random variables;
  \item the coordinates $X_j$ are linked by linear relations over
  $\mathbb{F}_2$ that survive in the limit;
  \item the discrete nature of the highest-bit map introduces local
  arithmetic correlations whose cumulative effect is difficult to
  quantify.
\end{itemize}
Nevertheless the Brownian-bridge prediction organizes the numerical
evidence for the \emph{height} of $\max_r\Delta(r)$ remarkably well and
supplies a concrete target for future analytic work. It does not
organize the recorded horizons.

\section{Two readings of the starting offset}
\label{sec:apps}

The range $\Delta(r)$ is a purely combinatorial statistic of the
degree sequence. It becomes a design quantity as soon as the starting
index $t$ is given a meaning outside the cycle. Two such meanings are
the offset of a pairing fragment and a change to the plaintext.

\subsection{Serial-knockout scheduling}

Recent work on the structural
stability of serial-knockout competitions
(SKCs)~\cite{LambersPendavinghSpieksma2025,
DevriesereCsatoGoossens2025} has emphasized that the choice of bracket
can materially affect fairness; the present square-root law quantifies
one concrete source of that sensitivity when the bracket is generated
from a contiguous $m$-sequence fragment. Suppose $N=2^k$ players are to
be placed into a single-elimination bracket that is generated from a
contiguous fragment of an $m$-sequence of period $M=2^k-1$. The starting
offset of the fragment determines the entire pairing sequence: which
players meet in the first round, which winners advance along which
paths, and ultimately how balanced the bracket is with respect to a
fixed seeding or strength ordering.

Identify player strength classes with the degree values
$0,1,\ldots,k-1$. The empirical histogram of a window of length $r$
then becomes the histogram of opponent classes encountered along a
typical path of length $r$. The degree discrepancy $D(t,r)$ then
measures the deviation of that path from the ideally balanced
distribution $q_j=2^j/M$. Consequently the range
\[
\Delta(r)=\max_t D(t,r)-\min_t D(t,r)
\]
is precisely the gap between the most balanced and the least balanced
choice of starting offset. The square-root law therefore supplies a
quantitative answer to a practical question:
\begin{quote}
\emph{How much fairness can be lost by an unfortunate choice of the
initial offset?}
\end{quote}
For a competition whose relevant horizon is a positive fraction of the
full period, the answer is of order $\sqrt{N}=2^{k/2}$. In other words, the
worst-case imbalance grows like the square root of the number of
players, independently of any further arithmetic fine structure of the
field. When the organizer is able to search for a good offset the same
law shows that the benefit of that search is also of order $2^{k/2}$:
the difference between the best and the worst schedule is not
negligible, yet it remains far smaller than the linear worst-case
bounds that would hold for arbitrary (non-algebraic) designs.

The location data of Section~\ref{sec:location} add a design caveat.
If the recorded maximizer of $\Delta(r)$ can sit at a short horizon,
then an organizer who only checks mid-season balance may miss the
worst offset sensitivity. Complement symmetry~\eqref{eq:complement}
still implies that a short horizon and its complementary long horizon
have the same range, so it is enough to inspect $\rho\le 1/2$; it is
not enough to inspect only a neighbourhood of $\rho=1/2$.

In the short-horizon regime $r\ll N$ that often arises in practice
the same picture persists, only with a smaller implicit constant. The
profiles already show substantial range at moderate relative lengths,
so the square-root prediction remains relevant even when the bracket
depth is a modest fraction of the theoretical maximum.

As a concrete illustration, for a $32$-player tournament ($k=5$) the
observed peak is already $\Delta\approx 2.3$, while for a $1024$-player
tournament ($k=10$) it is $\Delta\approx 22.9$. These gaps are large
enough to be felt by participants, yet they remain far smaller than the
linear discrepancies that can appear in non-algebraic constructions.

Thus the empirical law is not merely a statement about $m$-sequences;
it is a concrete design guideline for any SKC that draws its pairings
from a linear-feedback construction.

\subsection{A cryptographic reading}

The same index $t$ is a change to the plaintext: a nonzero difference
selects a starting point on the multiplicative orbit, and the degree
of each successive state is the degree of the corresponding output
difference~\cite{GolombGong2005,Paakkonen2026DAM}. Then $\Delta(r)$ is
the gap between plaintext differences whose degree histograms are most
and least balanced. This is only a reading of the statistic already
computed above, not an analysis of a cipher.

\section{A conjecture and open questions}
\label{sec:open}

The profiles up to $k=28$ make a square-root law for the
\emph{height} of $\max_r\Delta(r)$ the natural target. We state that
target as a conjecture. The recorded location of the maximizer is
deliberately omitted: the grids do not support a mid-path claim.

\begin{conjecture}[Square-root law]
\label{conj:sqrt}
There exist absolute constants $0<c_1\le c_2<\infty$ such that
\[
c_1\cdot 2^{k/2}
\le
\max_{1\le r\le 2^k-1}\Delta(r)
\le
c_2\cdot 2^{k/2}
\]
for every $k\ge 2$. Moreover,
\[
\liminf_{k\to\infty}
\frac{\max_{1\le r\le 2^k-1}\Delta(r)}{2^{k/2}}
=\frac12
\qquad\text{and}\qquad
\limsup_{k\to\infty}
\frac{\max_{1\le r\le 2^k-1}\Delta(r)}{2^{k/2}}
\le 1.
\]
\end{conjecture}

The first display is a uniform $\Theta(2^{k/2})$ bound. The smallest
recorded $d_k$ is $d_2\approx 0.17$, so any $c_1\le 1/6$ already
matches the computed range; the interesting constant is the
asymptotic floor $1/2$. The upper envelope $c_2=1$ is a conjecture
about the exhaustive maximum. The grid values $d_k\le 1$ do not imply
it, because a finer mesh could still overshoot. The free-walk
heuristic and the absence of any overshoot on $R_k$ are the reasons
for writing $\le 1$ rather than a larger $c_2$. We do not claim that
the limsup equals $1$; it could be strictly smaller (for instance
$1/\sqrt{2}$).

\begin{quote}
\textbf{Question 1.}
Prove Conjecture~\ref{conj:sqrt}. Even the weaker lower bound
$\max_r\Delta(r)=\Omega\bigl(2^{k/2}/\mathrm{polylog}\,k\bigr)$
would be a first step. A finite-check lower bound is already
available: the profile peaks of Table~\ref{tab:peaks} are lower
bounds on $\max_r\Delta(r)$, and for $k=2,3,4$ the grid already
includes every $r$.
\end{quote}

\begin{quote}
\textbf{Question 2.}
If the limits in Conjecture~\ref{conj:sqrt} exist, are the constants
sharp? In particular, is the limsup equal to $1$, or strictly smaller
than $1$? No computed $d_k$ exceeds $1$, and the closest approach is
$d_{15}\approx 0.98$.
\end{quote}

\begin{quote}
\textbf{Question 3.}
How does the range $\max_r\Delta(r)$ relate to the fixed-horizon
total-variation gaps $D_{\min}(k,r)$ and $D_{\max}(k,r)$ studied
in~\cite{Paakkonen2026DAM}? In particular, for which regimes of the
relative length $\rho=r/M$ is the $\ell_1$-range essentially determined
by the extremal TV values at that horizon? Complement
symmetry~\eqref{eq:complement} reduces the question to $\rho\le 1/2$.
\end{quote}

\begin{quote}
\textbf{Question 4.}
The $28$-point profiles do not place the first recorded maximizer in a
fixed bulk interval (Section~\ref{sec:location}). Is the
\emph{exhaustive} maximizer still off-centre for those $k$ whose grid
argmax is short, or does a finer grid pull $\rho^\ast$ back to a bulk
interval that the coarse mesh sampled unevenly? A finite check for one
mid-size exact $k$ (for instance $k=12$ or $k=14$) would already
separate a genuine short-window peak from a grid artefact.
Complement symmetry~\eqref{eq:complement} again reduces the search to
$\rho\le 1/2$.
\end{quote}

\begin{quote}
\textbf{Question 5.}
How strongly does the oscillation of the normalized sequence $d_k$
depend on the arithmetic nature of $M=2^k-1$? In particular, is the
distribution of peaks systematically different when $M$ is a Mersenne
prime from when $M$ is highly composite (or, more refinedly, when the
number of divisors of $M$ is large)? The present data set is too short
to decide this. The same question can now be asked of the folded
location $\min(\rho_k^\ast,1-\rho_k^\ast)$.
\end{quote}

A genuine invariance principle that would upgrade the Brownian-bridge
heuristic to a proof of Conjecture~\ref{conj:sqrt} still seems distant:
the degree sequence is deterministic, algebraically rigid, and
constrained by linear relations over $\mathbb{F}_2$. The conjecture is
a statement about height only. The location of the maximizer remains a
separate experimental question.

\section*{Data and code availability}
The Python code used to generate the degree sequences and compute the
sliding-window ranges $\Delta(r)$, together with the full degree
discrepancy profiles for $k=2,\dots,28$ and the scripts that produce
Figures~\ref{fig:dk} and~\ref{fig:argmax}, is available at
\begin{center}
\small
\url{https://github.com/fv2fm2z7rk-droid/An-Empirical-Square-Root-Law-for-the-Degree-Discrepancy-of-Binary-m-Sequences}
\end{center}

\section*{Declaration of generative AI and AI-assisted technologies
in the writing process}

During the preparation of this work, the author relied primarily on
Cursor Grok 4.6 as the main AI assistant, supplemented by OpenAI
Codex, and Google's Gemini. These
tools were used to assist with language editing, \LaTeX\ formatting,
exploratory discussion of related ideas, generation and refinement of
Python code, extraction of recorded argmaxes from the existing
profiles, and manuscript consistency checks against the computational
profiles. The AI tools were used as assistants only. The central
research idea, the computational results, the mathematical arguments,
and all intellectual content are the author's own. The author reviewed,
verified, and edited all AI-assisted outputs and takes full
responsibility for the content of the submitted manuscript.

\section*{Funding}

This research received no external funding.

\section*{Disclosure statement}

The author reports there are no competing interests to declare.

\end{document}